\documentclass[11pt]{article}
\usepackage{graphicx}
\usepackage{epsfig}
\usepackage{epsf}
\usepackage{epstopdf}

\begin{document}
\pagestyle{empty}

%

\newcommand{\bea}{\begin{eqnarray}}  
\newcommand{\eea}{\end{eqnarray}}  
\def\eqa{&=&}  
\def\ccr{\nonumber\\}  
  
\def\a{\alpha}
\def\b{\beta}
\def\m{\mu}
\def\n{\nu}
\def\r{\rho}
\def\s{\sigma}
\def\ep{\epsilon}

\def\cosech{\rm cosech}
\def\sech{\rm sech}
\def\coth{\rm coth}
\def\tanh{\rm tanh}

\def\sqr#1#2{{\vcenter{\vbox{\hrule height.#2pt  
     \hbox{\vrule width.#2pt height#1pt \kern#1pt  
           \vrule width.#2pt}  
       \hrule height.#2pt}}}}  
\def\square{\mathchoice\sqr66\sqr66\sqr{2.1}3\sqr{1.5}3}  
  
\def\appendix{\par\clearpage
  \setcounter{section}{0}
  \setcounter{subsection}{0}
  \def\@sectname{Appendix~}
  \def\theequation{\thesection\arabic{equation}}
  \def\thesection{\Alph{section}}}
 
\def\thefigures#1{\par\clearpage\section*{Figures\@mkboth
  {FIGURES}{FIGURES}}\list
  {Fig.~\arabic{enumi}.}{\labelwidth\parindent\advance
\labelwidth -\labelsep
      \leftmargin\parindent\usecounter{enumi}}}
\def\figitem#1{\item\label{#1}}
\let\endthefigures=\endlist
 
\def\thetables#1{\par\clearpage\section*{Tables\@mkboth
  {TABLES}{TABLES}}\list
  {Table~\Roman{enumi}.}{\labelwidth-\labelsep
      \leftmargin0pt\usecounter{enumi}}}
\def\tableitem#1{\item\label{#1}}
\let\endthetables=\endlist
 
\def\@sect#1#2#3#4#5#6[#7]#8{\ifnum #2>\c@secnumdepth
     \def\@svsec{}\else
     \refstepcounter{#1}\edef\@svsec{\@sectname\csname the#1\endcsname
.\hskip 1em }\fi
     \@tempskipa #5\relax
      \ifdim \@tempskipa>\z@
        \begingroup #6\relax
          \@hangfrom{\hskip #3\relax\@svsec}{\interlinepenalty \@M #8\par}
        \endgroup
       \csname #1mark\endcsname{#7}\addcontentsline
         {toc}{#1}{\ifnum #2>\c@secnumdepth \else
                      \protect\numberline{\csname the#1\endcsname}\fi
                    #7}\else
        \def\@svse=chd{#6\hskip #3\@svsec #8\csname #1mark\endcsname
                      {#7}\addcontentsline
                           {toc}{#1}{\ifnum #2>\c@secnumdepth \else
                             \protect\numberline{\csname the#1\endcsname}\fi
                       #7}}\fi
     \@xsect{#5}}
 
\def\@sectname{}
%
%
\def\eg{\hbox{\it e.g.}}        \def\cf{\hbox{\it cf.}}
\def\etal{\hbox{\it et al.}}
\def\dash{\hbox{---}}
\def\bR{\mathop{\bf R}}
\def\bC{\mathop{\bf C}}
\def\eq#1{{eq. \ref{#1}}}
\def\eqs#1#2{{eqs. \ref{#1}--\ref{#2}}}
\def\lie{\hbox{\it \$}} 
\def\partder#1#2{{\partial #1\over\partial #2}}
\def\secder#1#2#3{{\partial^2 #1\over\partial #2 \partial #3}}
\def\abs#1{\left| #1\right|}
\def\ltap{\ \raisebox{-.4ex}{\rlap{$\sim$}} \raisebox{.4ex}{$<$}\ }
\def\gtap{\ \raisebox{-.4ex}{\rlap{$\sim$}} \raisebox{.4ex}{$>$}\ }
\def\contract{\makebox[1.2em][c]{
        \mbox{\rule{.6em}{.01truein}\rule{.01truein}{.6em}}}}
%
\def\com#1#2{
        \left[#1, #2\right]}
%
%
\def\bentarrow{\:\raisebox{1.3ex}{\rlap{$\vert$}}\!\rightarrow}
\def\longbent{\:\raisebox{3.5ex}{\rlap{$\vert$}}\raisebox{1.3ex}%
        {\rlap{$\vert$}}\!\rightarrow}
\def\onedk#1#2{
        \begin{equation}
        \begin{array}{l}
         #1 \\
         \bentarrow #2
        \end{array}
        \end{equation}
                }
\def\dk#1#2#3{
        \begin{equation}
        \begin{array}{r c l}
        #1 & \rightarrow & #2 \\
         & & \bentarrow #3
        \end{array}
        \end{equation}
                }
\def\dkp#1#2#3#4{
        \begin{equation}
        \begin{array}{r c l}
        #1 & \rightarrow & #2#3 \\
         & & \phantom{\; #2}\bentarrow #4
        \end{array}
        \end{equation}
                }
\def\bothdk#1#2#3#4#5{
        \begin{equation}
        \begin{array}{r c l}
        #1 & \rightarrow & #2#3 \\
         & & \:\raisebox{1.3ex}{\rlap{$\vert$}}\raisebox{-0.5ex}{$\vert$}%
        \phantom{#2}\!\bentarrow #4 \\
         & & \bentarrow #5
        \end{array}
        \end{equation}
                }
\newcommand{\nc}{\newcommand}
\nc{\spa}[3]{\left\langle#1\,#3\right\rangle}
\nc{\spb}[3]{\left[#1\,#3\right]}
\nc{\ksl}{\not{\hbox{\kern-2.3pt $k$}}}
\nc{\hf}{\textstyle{1\over2}}
\nc{\pol}{\varepsilon}
\nc{\tq}{{\tilde q}}
\nc{\esl}{\not{\hbox{\kern-2.3pt $\pol$}}}
\newcommand{\1}{{\'\i}}
\newcommand{\be}{\begin{equation}}
\newcommand{\ee}{\end{equation}\noindent}
\newcommand{\bear}{\begin{eqnarray}}
\newcommand{\ear}{\end{eqnarray}\noindent}
\newcommand{\benn}{\begin{enumerate}}
\newcommand{\enn}{\end{enumerate}}
\newcommand{\no}{\noindent}
\date{}
\renewcommand{\theequation}{\arabic{section}.\arabic{equation}}
\renewcommand{\arraystretch}{2.5}
\newcommand{\GeV}{\mbox{GeV}}
\newcommand{\cL}{\cal L}
\newcommand{\D}{\cal D}
\newcommand{\Dhalf}{{D\over 2}}
\newcommand{\Det}{{\rm Det}}
\newcommand{\PP}{\cal P}
\newcommand{\G}{{\cal G}}
\def\R{1\!\!{\rm R}}
\def\Eins{\mathord{1\hskip -1.5pt
\vrule width .5pt height 7.75pt depth -.2pt \hskip -1.2pt
\vrule width 2.5pt height .3pt depth -.05pt \hskip 1.5pt}}
\newcommand{\symb}{\mbox{symb}}
\renewcommand{\arraystretch}{2.5}
\newcommand{\slD}{\raise.15ex\hbox{$/$}\kern-.57em\hbox{$D$}}
\newcommand{\slpartial}{\raise.15ex\hbox{$/$}\kern-.57em\hbox{$\partial$}}
\newcommand{\slG}{{{\dot G}\!\!\!\! \raise.15ex\hbox {/}}}
\newcommand{\Gd}{{\dot G}}
\newcommand{\Gund}{{\underline{\dot G}}}
\newcommand{\Gdd}{{\ddot G}}
\def\GBd12{{\dot G}_{B12}}
\def\mneg{\!\!\!\!\!\!\!\!\!\!}
\def\Mneg{\!\!\!\!\!\!\!\!\!\!\!\!\!\!\!\!\!\!\!\!}
\def\non{\nonumber}
\def\beqn*{\begin{eqnarray*}}
\def\eqn*{\end{eqnarray*}}
\def\sy{\scriptscriptstyle}
\def\footstrut{\baselineskip 12pt}
\def\square{\kern1pt\vbox{\hrule height 1.2pt\hbox{\vrule width 1.2pt
   \hskip 3pt\vbox{\vskip 6pt}\hskip 3pt\vrule width 0.6pt}
   \hrule height 0.6pt}\kern1pt}
\def\np{n_{+}}
\def\nm{n_{-}}
\def\Np{N_{+}}
\def\Nm{N_{-}}
\def\exmn{\Bigl(\mu \leftrightarrow \nu \Bigr)}
\def\slash#1{#1\!\!\!\raise.15ex\hbox {/}}
\def\dint#1{\int\!\!\!\!\!\int\limits_{\!\!#1}}
\def\bra#1{\langle #1 |}
\def\ket#1{| #1 \rangle}
\def\vev#1{\langle #1 \rangle}
\def\rightvac{\mid 0\rangle}
\def\leftvac{\langle 0\mid}
\def\dps{\displaystyle}
\def\sy{\scriptscriptstyle}
\def\half{{1\over 2}}
\def\third{{1\over3}}
\def\fourth{{1\over4}}
\def\fifth{{1\over5}}
\def\sixth{{1\over6}}
\def\seventh{{1\over7}}
\def\eigth{{1\over8}}
\def\ninth{{1\over9}}
\def\tenth{{1\over10}}
\def\pa{\partial}
\def\ddtau{{d\over d\tau}}
\def\ge{\hbox{\textfont1=\tame $\gamma_1$}}
\def\gz{\hbox{\textfont1=\tame $\gamma_2$}}
\def\gd{\hbox{\textfont1=\tame $\gamma_3$}}
\def\go{\hbox{\textfont1=\tamt $\gamma_1$}}
\def\gt{\hbox{\textfont1=\tamt $\gamma_2$}}
\def\gth{\hbox{\textfont1=\tamt $\gamma_3$}} 
\def\gf{\hbox{$\gamma_5\;$}}
\def\ie{\hbox{$\textstyle{\int_1}$}}
\def\iz{\hbox{$\textstyle{\int_2}$}}
\def\id{\hbox{$\textstyle{\int_3}$}}
\def\ldop{\hbox{$\lbrace\mskip -4.5mu\mid$}}
\def\rdop{\hbox{$\mid\mskip -4.3mu\rbrace$}}
\def\eps{\epsilon}
\def\epshalf{{\epsilon\over 2}}
\def\e{\mbox{e}}
\def\mn{{\mu\nu}}
\def\exmn{{(\mu\leftrightarrow\nu )}}
\def\ab{{\alpha\beta}}
\def\exab{{(\alpha\leftrightarrow\beta )}}
\def\g{\mbox{g}}
\def\kinb{{1\over 4}\dot x^2}
\def\kinf{{1\over 2}\psi\dot\psi}
\def\expk{{\rm exp}\biggl[\,\sum_{i<j=1}^4 G_{Bij}k_i\cdot k_j\biggr]}
\def\expp{{\rm exp}\biggl[\,\sum_{i<j=1}^4 G_{Bij}p_i\cdot p_j\biggr]}
\def\expshort{{\e}^{\half G_{Bij}k_i\cdot k_j}}
\def\expabb{{\e}^{(\cdot )}}
\def\epseps#1#2{\varepsilon_{#1}\cdot \varepsilon_{#2}}
\def\epsk#1#2{\varepsilon_{#1}\cdot k_{#2}}
\def\kk#1#2{k_{#1}\cdot k_{#2}}
\def\G#1#2{G_{B#1#2}}
\def\Gp#1#2{{\dot G_{B#1#2}}}
\def\GF#1#2{G_{F#1#2}}
\def\Dab{{(x_a-x_b)}}
\def\Dsq{{({(x_a-x_b)}^2)}}
\def\lag{( -\partial^2 + V)}
\def\PITD{{(4\pi T)}^{-{D\over 2}}}
\def\4piTD{{(4\pi T)}^{-{D\over 2}}}
\def\4piT4{{(4\pi T)}^{-2}}
\def\TintmD{{\dps\int_{0}^{\infty}}{dT\over T}\,e^{-m^2T}
    {(4\pi T)}^{-{D\over 2}}}
\def\Tintm4{{\dps\int_{0}^{\infty}}{dT\over T}\,e^{-m^2T}
    {(4\pi T)}^{-2}}
\def\Tintm{{\dps\int_{0}^{\infty}}{dT\over T}\,e^{-m^2T}}
\def\Tint{{\dps\int_{0}^{\infty}}{dT\over T}}
\def\pint{{\dps\int}{dp_i\over {(2\pi)}^d}}
\def\Dx{\dps\int{\cal D}x}
\def\Dy{\dps\int{\cal D}y}
\def\Dpsi{\dps\int{\cal D}\psi}
\def\Tr{{\rm Tr}\,}
\def\tr{{\rm tr}\,}
\def\sumij{\sum_{i<j}}
\def\freeexp{{\rm e}^{-\int_0^Td\tau {1\over 4}\dot x^2}}
\def\arraystretch{2.5}
\def\Ge{\mbox{GeV}}
\def\dA{\partial^2}
\def\DA{\sqsubset\!\!\!\!\sqsupset}
\def\FFdual{F\cdot\tilde F}
\def\mn{{\mu\nu}}
\def\rs{{\rho\sigma}}
\def\oplusotimes{{{\lower 15pt\hbox{$\scriptscriptstyle \oplus$}}\atop{\otimes}}}
\def\perppar{{{\lower 15pt\hbox{$\scriptscriptstyle \perp$}}\atop{\parallel}}}
\def\oopp{{{\lower 15pt\hbox{$\scriptscriptstyle \oplus$}}\atop{\otimes}}\!{{\lower 15pt\hbox{$\scriptscriptstyle \perp$}}\atop{\parallel}}}
%
%
\def\bbbr{{\rm I\!R}}
\def\bbbone{{\mathchoice {\rm 1\mskip-4mu l} {\rm 1\mskip-4mu l}
{\rm 1\mskip-4.5mu l} {\rm 1\mskip-5mu l}}}
\def\bbbz{{\mathchoice {\hbox{$\sf\textstyle Z\kern-0.4em Z$}}
{\hbox{$\sf\textstyle Z\kern-0.4em Z$}}
{\hbox{$\sf\scriptstyle Z\kern-0.3em Z$}}
{\hbox{$\sf\scriptscriptstyle Z\kern-0.2em Z$}}}}

\renewcommand{\thefootnote}{\protect\arabic{footnote}}
\hfill {\large AEI-2005-137}

\begin{center}
{\huge\bf One loop photon-graviton mixing in an electromagnetic field: Part 2}
\vskip1.3cm
{\large F. Bastianelli$^{a,b}$, U. Nucamendi$^{c}$, 
C. Schubert$^{a,c}$, V. M. Villanueva$^{a,c}$}
\\[1.5ex]

\begin{itemize}
\item [$^a$]
{\it 
Max-Planck-Institut f\"ur Gravitationsphysik, Albert-Einstein-Institut,
M\"uhlenberg 1, D-14476 Potsdam, Germany
}
\item [$^b$]
{\it
Dipartimento di Fisica, Universit\`a di Bologna and INFN, Sezione di Bologna,
Via Irnerio 46, I-40126 Bologna, Italy
}
\item [$^c$]
{\it 
Instituto de F\'{\i}sica y Matem\'aticas
\\
Universidad Michoacana de San Nicol\'as de Hidalgo\\
Edificio C-3, Apdo. Postal 2-82\\
C.P. 58040, Morelia, Michoac\'an, M\'exico\\
}
\end{itemize}
\vspace{1cm}
 {\large \bf Abstract}
\end{center}
\begin{quotation}
In part 1 of this series compact
integral representations had been obtained for the one-loop
photon-graviton amplitude involving a charged spin 0 or spin 1/2 
particle in the loop and an arbitrary constant electromagnetic field.
In this sequel, we study the structure and magnitude of the various polarization
components of this amplitude on-shell.
Explicit expressions are obtained for a number of
limiting cases. 
 
\end{quotation}
\vfill\eject
\pagestyle{plain}
\setcounter{page}{1}
\setcounter{footnote}{0}

\vspace{10pt}
\section{Introduction}
\renewcommand{\theequation}{1.\arabic{equation}}
\setcounter{equation}{0}

As has been recognized 
many years ago \cite{gertsenshtein,zelnovbook} the 
quantized Einstein-Maxwell theory predicts
the process of photon-graviton conversion in
an electromagnetic field. 
The tree level vertex for this amplitude is (see appendix A)

\bear
\half
\kappa 
h_\mn\Bigl(F^{\mu\alpha}f^{\nu}_{\,\,\,\alpha} + f^{\mu}_{\,\,\alpha}\,F^{\nu\alpha}
\Bigr) - \fourth\kappa h^{\mu}_{\mu}F^{\ab} f_\ab.
\label{treelevel}
\ear
Here $h_{\mu\nu}$ denotes the graviton, $f_{\mu\nu}$ the photon, and $F^{\mu\nu}$
the external field. $\kappa$ is the gravitational coupling constant.
The corresponding photon-graviton vertex in momentum space reads
$-{i\over 2}\kappa C^{\mn,\alpha}$, where

\bear
C^{\mn,\alpha} &=&
\bigl(F\cdot k\bigr)^{\alpha}\eta^{\mn}
+F^{\mu\alpha}k^{\nu}
+F^{\nu\alpha}k^{\mu}
-\bigl(F\cdot k\bigr)^{\mu}\eta^{\nu\alpha}
-\bigl(F\cdot k\bigr)^{\nu}\eta^{\mu\alpha} \, .
\non\\&&
\label{defCmna}
\ear
This interaction leads, assuming sufficient coherence of propagation,
to photon-graviton oscillations which are analogous
to the better-known neutrino flavour \cite{bilpon} and 
photon-axion oscillations \cite{primakoff,dktw,rafsto}.
The true eigenstates of propagation in a background
field will in general be certain mixtures of photon and
graviton states. Determining these eigenstates and their
dispersion relations requires, at tree level and in Fourier space, the diagonalization 
of the following matrix
(see eqs. (\ref{eom}), (\ref{Pitree}), and (\ref{Pigravtree})):

\bear
\hspace{-20pt}
\Biggl(
\matrix {\eta^{\alpha\beta}k^2-k^{\alpha}k^{\beta}&
{i\over 2}\kappa C^{\kappa\lambda,\alpha} 
\cr
{i\over 2}\kappa C^{\mn,\beta}&
{k^2\over 4}(\eta^{\mu\kappa}\eta^{\nu\lambda}
\!+\!
\eta^{\mu\lambda}\eta^{\nu\kappa}
\!-\!2\eta^{\mn}\eta^{\kappa\lambda} + \! {\scriptstyle \ldots}) 
\! \cr
}
\Biggr) 
\Bigl(
\matrix {a_{\beta}(k) \cr
h_{\kappa\lambda}(k) \cr
}
\Bigr)  
&& = 0\, .\non\\
\label{disptree}
\ear
Here $a_{\beta}(k)$ represents the photon and $h_{\kappa\lambda}(k)$ the
graviton.
For many cases of physical interest, this problem can be
simplified assuming the field to be homogeneous or near-homogeneous, 
and the modified dispersion relations to be close to the vacuum
ones. An efficient formalism for calculating the evolution of the
photon-graviton or photon-axion 
system under these conditions was developed
in \cite{rafsto}. 
 
Due to the smallness of the gravitational coupling $\kappa$, the
photon-graviton mixing case has received less attention than the
photon-axion one. 
Nevertheless, a number of authors have studied
possible observable effects \cite{zelnovbook,rafsto,losotr,magueijo,chen,cilhar,raffeltbook}.  
The chances of observing this process in the laboratory appear very remote.
As with other processes involving very small
couplings, the natural setting is astrophysics where one can hope to at least  partially
compensate this smallness by large field strengths or exposure times.
In principle, any process based on photon-axion conversion in a field 
(see, e.g., \cite{sikivie,morris,cskate,grrozu,mobego,ckpt,baskun}) must 
have an analogue based on photon-graviton conversion. 
In \cite{rafsto} photon-graviton conversion was considered in a pulsar field, as well as
in the galactic magnetic field, but the effect was found to be very small.
Photon-graviton conversion in a primordial magnetic
field has been proposed as a possible contribution to the
cosmic microwave background anisotropy \cite{magueijo,chen}.
However, taking plasma effects into account renders the effect negligible
\cite{cilhar}. 

A natural enhancement of the photon-graviton oscillation occurs in
theories with extra dimensions \cite{ardidv} due to the existence of
an infinite tower of Kaluza-Klein gravitons. 
In \cite{defuza} both the effect of the photon-graviton oscillation on the cosmic microwave background and the conversion in a pulsar background were reconsidered in this context, but the
enhancement was found to be insufficient to lift these effects into the observable range. 
  
To our knowledge, the photon-graviton process has previously been studied
only at the tree level. In the first part of this series \cite{part1} (referred to as `part 1' in the following)
we considered the one-loop corrections to this amplitude due to massive charged
spin 0 (denoted ${\bar \Pi}^{\mu\nu ,\alpha}_{\rm scal}$)
and spin 1/2 particles (denoted ${\bar \Pi}^{\mu\nu ,\alpha}_{\rm spin}$) in the loop (fig. 1).

\begin{figure}[h]
\vspace{50pt}
\centering
\includegraphics{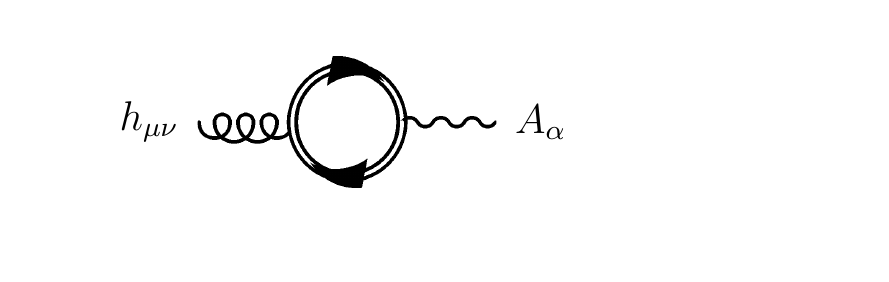}
\caption{One-loop photon-graviton amplitude in a constant field. The double line represents
the propagator of a charged scalar or spin $\half$ particle in a constant field.}
\label{fig1}
\end{figure}
 
 \vspace{20pt}
 
\medskip
Our motivation for considering this loop correction is twofold.  First, 
the dispersion relation (\ref{disptree}) will 
be modified in the following way by the one-loop contributions
(see eqs. (\ref{eom}), (\ref{Pitree}), and (\ref{Pigravtree})):

\bear
&&
\hspace{-20pt}
\biggl(
\matrix {\eta^{\alpha\beta}k^2-k^{\alpha}k^{\beta}-
\bar\Pi^{\alpha,\beta}&
{i\over 2}\kappa C^{\kappa\lambda,\alpha} - \bar\Pi^{\kappa\lambda,\alpha}
\cr
{i\over 2}\kappa C^{\mn,\beta}-\bar\Pi^{\mn,\beta}&
{k^2\over 4}(\eta^{\mu\kappa}\eta^{\nu\lambda}
\!+\!
\eta^{\mu\lambda}\eta^{\nu\kappa}
\!-\!2\eta^{\mn}\eta^{\kappa\lambda} + \! {\scriptstyle \ldots}) 
\!-\!\bar\Pi^{\mn,\kappa\lambda} \cr
}
\biggr) 
\Bigl(
\matrix {a_{\beta}(k) \cr
h_{\kappa\lambda}(k) \cr
}
\Bigr)   \non\\
&&\non\\
&& = 0\, .
\label{dispbig}
\ear
This equation involves the full one-loop photon-photon, photon-graviton
and graviton-graviton amplitudes, computed in the constant external field,
summed over all possible loop particles
(including, e.g., photons and gluons for the graviton propagator).
As is well-known, already in the pure QED case the one-loop
corrected dispersion relation in an external field

\bear
\Bigl[\eta^{\mn}k^2-k^{\mu}k^{\nu}-\Pi^{\mu,\nu}(k)\Bigr]a_{\nu}(k) &=& 0
\label{dispmod}
\ear
leads to highly nontrivial deviations from the vacuum case 
\cite{toll,minguzzi,baibre,biabia,adler,batsha,ritus_annphys,tsaerb,bakast,melsto,ditgiebook,kohyam}. 
We believe that it will be very instructive to generalize this study to the
mixed photon-graviton system (\ref{dispbig}), particularly considering
the fact that in curved backgrounds superluminal phase velocities
are known to occur \cite{druhat,lapata,shore}. See also \cite{holsho}
for recent studies
on the photon vacuum polarization in curved space
at arbitrary frequencies and related investigations on a
possible breakdown of microcausality.

Second, while the tree level interaction term (\ref{treelevel}) depends linearly on the background field,
the one-loop corrections depend nontrivially on the field strength as well as on the photon/graviton energy. Thus, although at linear order in $F_{\mn}$ the one-loop corrections are 
down by an explicit factor of $\alpha$ compared to the tree level term, it is a priori
conceivable that for sufficiently large fields and some range of photon energies the 
one-loop amplitudes would dominate over the tree level one. 

The plan of this paper is as follows. In section \ref{polarize} we 
find choices of physical polarizations well-adapted to the
structure of the worldline parameter integrals obtained in part 1. 
In section \ref{onshell} we specialize to the on-shell case,
and introduce some convenient notation. In section \ref{magnetic}
we consider the purely magnetic case, and
explicitly evaluate the amplitude for three ranges of parameters:
in \ref{belowthreshold}  we present a direct numerical evaluation
for arbitrary field strength and photon/graviton energies below threshold;
in \ref{lowphotonmag} closed-form results are found for the 
zero energy limits; in \ref{weakfieldmag} we consider the case
of a weak field but arbitrary energies.

It seems that photon-graviton conversion so far has been studied 
only for the magnetic field case. Although the physical relevance
of this process is even more hypothetical in the electric field case,
in part 1 we kept the electric field component since, quite generally, 
in the worldline formalism calculations in a general
electromagnetic field are not substantially more difficult than in a purely magnetic field 
\cite{vv,review}.
In this sequel, too, we shortly consider the electric field case 
in section \ref{electric}, and present results for two cases where the electric
result can easily be inferred from the magnetic one, namely the
zero energy and weak field cases.
We summarize our results in section \ref{conclusions}.

\vspace{10pt}
\section{Polarization decomposition of the amplitude}
\label{polarize}
\renewcommand{\theequation}{2.\arabic{equation}}
\setcounter{equation}{0}

Let us now project the photon-graviton amplitude on physical polarizations.
We need to choose two photon polarization vectors $\varepsilon_{1,2}$
such that

\bear
\varepsilon_i^{\mu}\varepsilon_{j\mu} &=& \delta_{ij}, 
\quad
k_{\mu}\varepsilon_i^{\mu} = 0 \, .
\label{condphotpol}
\ear
Similarly, we need two symmetric, traceless, and transverse graviton polarization tensors $\varepsilon_{1,2}^{\mn}$:

\bear
\varepsilon_{i}^{\mn}\varepsilon_{j\mn} = 2\delta_{ij},\quad\quad
\varepsilon_i^{\mu\nu} =\varepsilon_i^{\nu\mu},\quad 
\quad \varepsilon^{\mu}_{i\mu} = 0, \quad \quad
k_{\mu}\varepsilon_i^{\mn} =0 \, .
\nonumber\\
\label{condgravpol}
\ear
As we will see, substantial simplifications can be achieved by choosing polarization
vectors which are adapted to the background field. In part 1 we
had, following \cite{vv,review}, 
written the worldline Green functions using the matrix basis 
$\hat{\cal Z}_{\pm}$, $\hat{\cal Z}_{\pm}^2$,

\bear
\hat{\cal Z}_+^\mn &=& {aF^\mn -b\tilde F^\mn\over a^2 + b^2} ,\qquad
\hat{\cal Z}_-^\mn = -i{bF^\mn+a\tilde F^\mn\over a^2+b^2},\label{Z}\non\\
&&\non\\
\Bigl(\hat{\cal Z}_+^2\Bigr)^{\mu\nu} &=& {F^{\mu\lambda}F_{\lambda}^{\,\,\nu}
-b^2\eta^\mn \over a^2+b^2},\qquad
\Bigl(\hat{\cal Z}_-^2\Bigr)^\mn = -{F^{\mu\lambda}F_{\lambda}^{\,\,\nu}+a^2\eta^\mn\over a^2+b^2}.
\non\\
\non\\
\label{ZZsquare}
\ear
Here $\tilde F_{\mu\nu}=\half \varepsilon_{\mu\nu\alpha\beta}F^{\alpha\beta}$
is the dual field strength tensor
\footnote{We work in Minkowski space with
$\eta^{\mn}={\rm diag}(-+++)$ and $\varepsilon^{0123}=1$.} 
and $a$, $b$ are related to the two standard Maxwell invariants by
$a^2-b^2=B^2-E^2$, $ab = {\bf E}\cdot {\bf B}$. 

\no
Note the following properties of this basis \cite{vv,review},

\bear
\hat{\cal Z}_{\pm}^3 &=& - \hat{\cal Z}_{\pm} ,  \label{Zcube}\\
\hat{\cal Z}_+\cdot \hat{\cal Z}_- &= &0 \, .
\label{Zorth}
\ear
We now use this basis
to define polarization vectors as follows
\footnote{For the photon polarizations this basis has been introduced in \cite{ditgiebook} 
with different conventions and notations
(in particular, our $b$ corresponds to $-b$ there).}
,

\bear
\varepsilon_{\pm}^{\mu} &\equiv & {\bigl(\hat{\cal Z}_{\pm}\cdot k\bigr)^{\mu}\over
\lambda_{\pm}} \, . \non\\
\label{defpolphot}
\ear
Here the $\lambda_{\pm}$'s are normalization factors,

\bear
\lambda_+ &\equiv&\sqrt{\bigl(\hat{\cal Z}_+\cdot k\bigr)\cdot
\bigl(\hat{\cal Z}_+\cdot k\bigr)}
 = \sqrt{- k\cdot \bigl(\hat{\cal Z}^2_+\bigr)\cdot k} \, ,\nonumber\\
 \lambda_- &\equiv&-i\sqrt{-\bigl(\hat{\cal Z}_-\cdot k\bigr)\cdot
\bigl(\hat{\cal Z}_-\cdot k\bigr)}
 = -i\sqrt{k\cdot \bigl(\hat{\cal Z}^2_-\bigr)\cdot k} \, .
\nonumber\\
\label{deflambda}
\ear
Together with the orthogonality relation 
(\ref{Zorth})
they ensure that
$\varepsilon_+,\varepsilon_-$ satisfy the conditions
(\ref{condphotpol}). 
Both $\lambda_{\pm}$ can vanish; in this case the corresponding polarization 
vector is lightlike and cannot be normalized. We also note that

\bear
\lambda_+^2 + \lambda_-^2 = k^2
\label{sumlambda}
\ear
(from (\ref{ZZsquare})). The explicit form of $\varepsilon_{\pm}$ in terms of
$\bf E$ and $\bf B$ is rather complicated in the general case.
However, as usual things simplify considerably if one specializes to a
Lorentz system where $\bf E$ and $\bf B$ are both pointing along the positive z - axis,
${\bf E} = (0,0,E), {\bf B} = (0,0,B)$. 
This implies that $a=B$ and $b=E$. Here the dependence of $\lambda_{\pm}$,
$\varepsilon_{\pm}$ on the field magnitudes drops out, leaving only a memory
of the field direction:

\bear
\lambda_+ &=& \sqrt{(k^1)^2+(k^2)^2} \equiv \lambda_{\perp},\qquad 
\lambda_- = -i\sqrt{(k^0)^2-(k^3)^2} \equiv \lambda_{\parallel}
\, , \non\\
\label{lambdaspecial}
\ear

\no
and

\bear
\varepsilon_+^{\mu} &=& {(0,k^2,-k^1,0)\over \sqrt{(k^1)^2+(k^2)^2}} \equiv \varepsilon_{\perp}^{\mu},\qquad
\varepsilon_-^{\mu} = {(k^3,0,0,k^0)\over \sqrt{(k^0)^2-(k^3)^2}}
\equiv \varepsilon_{\parallel}^{\mu} \, . \nonumber\\
\label{polspecial}
\ear
The subscripts $\perp,\parallel$ refer to the field direction.
Such a Lorentz system exists provided that ${\bf E}\cdot{\bf B} > 0$.  The
case ${\bf E}\cdot {\bf B}<0$ differs from this only by a parity transformation,
but the case ${\bf E}\cdot {\bf B} = 0$ needs to be considered separately:
This case can for $E>B$ be transformed into the purely electric and for $B>E$ into the
purely magnetic field case. In both cases $\lambda_{\pm}$ and $\varepsilon_{\pm}$ 
are the same as in (\ref{lambdaspecial}), (\ref{polspecial}) if the z-axis is chosen as the
field direction. The remaining possibility is that ${\bf E}\cdot {\bf B}=0$ and $E=B$, the
"crossed field" case; here $a=b=0$ and the above basis cannot be used.

To simplify further, without loss of generality we shall
assume that the photon propagation is in the $y-z$ plane ($k^1=0, k^2>0$).

It will be useful to construct also the graviton polarizations using the same building blocks.
We define

\bear
\varepsilon^{\oplus\mu\nu} &\equiv& \varepsilon^{+\mu}\varepsilon^{+\nu}
- \varepsilon^{-\mu}\varepsilon^{-\nu} \, ,\nonumber\\
\varepsilon^{{\otimes}\,\mu\nu} &\equiv& \varepsilon^{+\mu}\varepsilon^{-\nu}
+ \varepsilon^{-\mu}\varepsilon^{+\nu}  \, .\nonumber\\
\label{defpolgrav}
\ear
Then the conditions (\ref{condgravpol}) are fulfilled as a consequence of
(\ref{condphotpol}).

This basis is extremely convenient since, when contracting $\bar\Pi^{\mn,\alpha}_{\rm scal,spin}$
with these polarization vectors/tensors, many terms in the integral representations 
obtained in part 1 (eqs. (3.13) and (4.11) there)
drop out on account of the orthogonality relation (\ref{Zorth})
and the antisymmetry of $\hat{\cal Z}_{\pm}$. In particular, all terms involving
a factor of $\dot {\cal S}_{B12}\cdot k$ or $\ddot {\cal S}_{B12}\cdot k$ will vanish.

We remark that a more standard, 
but less convenient, basis would be obtained by removing
the time component of $\varepsilon_{\parallel}$ by a longitudinal shift,

\bear
\tilde\varepsilon_{\parallel}^{\mu} = \varepsilon_{\parallel}^{\mu}
- {k^3k^{\mu}\over k^0\sqrt{(k^0)^2-(k^3)^2}}
\label{shiftepsmin}
\ear
The equivalence
of these two choices may not seem obvious for the gravitational part,
since neither $C^{\mn,\alpha}$ nor $\bar\Pi^{\mn,\alpha}_{\rm scal,spin}$ are transversal
in the graviton indices. However, using the gravitational Ward identity ((A.13) of part 1) 
it is easily shown that the shift makes no difference for the matrix elements. 

\no
In the following, we will denote 

\bear
C^{Aa} &\equiv& \varepsilon_{\mn}^{A}C^{\mn,\alpha}
\varepsilon_{\alpha}^a
\label{defCAa}
\ear
etc. where $A = \oplus, \otimes$ and $a = \perp, \parallel$.
At the tree level, one obtains the following simple result:

\bear
C^{\oplus\perp}
&=&
-2B\lambda_{\perp} \, ,\nonumber\\
C^{\oplus\parallel}
&=&
2iE \lambda_{\parallel}\, , \nonumber\\
C^{\otimes\perp}
&=&
-2i E \lambda_{\parallel}  \, ,\nonumber\\
C^{\otimes\parallel} 
&=& -2B \lambda_{\perp} \, .\nonumber\\
\label{poldecomptree}
\ear
The fact that for a purely magnetic (electric) field $\varepsilon^{\oplus}$ couples only
to $\varepsilon^{\perp}$ ($\varepsilon^{\parallel}$) and  
$\varepsilon^{\otimes}$ only
to $\varepsilon^{\parallel}$ ($\varepsilon^{\perp}$) is a consequence of CP invariance
(see \cite{rafsto}). It must therefore also hold for the loop corrections.

For the one-loop correction, using the polarization choices (\ref{defpolphot}), 
(\ref{defpolgrav}) in (3.13) resp.  (4.11) of part 1 yields the following:

\bear
\bar\Pi_{\rm scal}^{Aa}
&=&
{e\kappa\over 64\pi^2}\int_0^{\infty}{ds\over s}
\e^{-ism^2}
\Biggl\lbrace
{z_+z_-\over\sinh(z_+)\sinh(z_-)}
\nonumber\\&&\hspace{50pt}\times
\int_0^1du\,
\e^{-is\Phi}
\tilde J_{\rm scal}^{Aa}+{2\over 3}ie C^{Aa}\Biggr\rbrace
\, , \nonumber\\
\bar\Pi_{\rm spin}^{Aa}
&=&
-{e\kappa\over 32\pi^2}\int_0^{\infty}{ds\over s}
\e^{-ism^2}
\Biggl\lbrace
{z_+z_-\over\tanh(z_+)\tanh(z_-)}
\nonumber\\&&\hspace{50pt}\times
\int_0^1du\,
\e^{-is\Phi}
\tilde J_{\rm spin}^{Aa}
-{4\over 3}ie C^{Aa}
\Biggr\rbrace
\, , \nonumber\\
\label{decomp}
\ear
where now $z_+=ieBs$, $z_-=-eEs$ and
(repeated indices $a,b,\ldots$ are to be summed
over $\pm$)

\bear
\Phi &=& -\half {\bar A^a\over z^a}\lambda_a^2 \, ,
\label{Phifin}
\ear

\bear
\tilde J_{\rm scal}^{\oplus \perp} &=& 
{2\over s}(z_+  A^+_{B11}-z_- A^-_{B11})\bar A^+_{B12}\lambda_+
\non\\&&
+{2\over s}z_+\Bigl[(S_{B12}^+)^2-\bar A_{B12}^+\Bigl(A_{B12}^+ +{1\over z_+}\Bigr)\Bigr]
\lambda_+
\non\\&&
+i\bar A_{B12}^+\lambda_+
\biggl[\Bigl((S_{B12}^+)^2-(\bar A_{B12}^+)^2\Bigr)\lambda_+^2
+\Bigl(S_{B12}^+S_{B12}^- +(\bar A_{B12}^-)^2\Bigr)\lambda_-^2\biggr] \, ,
\non\\
\non\\
\tilde J_{\rm scal}^{{\otimes}\perp} &=& 
{2\over s}\Bigl[
S_{B12}^+z_-S_{B12}^-
-\bar A_{B12}^-z_+\Bigl(A_{B12}^+ +{1\over z_+}\Bigr)
\Bigr]\lambda_-
\non\\&&
+i\bar A_{B12}^-\lambda_-
\Bigl[
S_{B12}^+ S_{B12}^a\lambda_a^2 -2(\bar A_{B12}^+)^2\lambda_+^2
\Bigr] \, ,
\non\\
\label{decompJscal}
\ear

\bear
\tilde J_{\rm spin}^{\oplus\perp} &=& 
{2\over s}\Bigl[z_+ (A^+_{B11}-A^+_{F11})-z_- 
(A^-_{B11}-A^-_{F11})\Bigr](\bar A^+_{B12}+A^+_{F22})\lambda_+
\non\\&&
+{2\over s}z_+\Bigl[(S_{B12}^+)^2-(S_{F12}^+)^2+(A_{F12}^+)^2
-(\bar A_{B12}^++A^+_{F11})\Bigl(A_{B12}^+ +{1\over z_+}\Bigr)\Bigr]
\lambda_+
\non\\&&
+i\bar A_{B12}^+\lambda_+
(S^+_{B12}S^a_{B12}-S^+_{F12}S^a_{F12})\lambda_a^2
\nonumber\\&&
-i\bar A^+_{B12}\Bigl[(\bar A^+_{B12}+A^+_{F11})^2-(A^+_{F12})^2\Bigr]
\lambda_+^3 \nonumber\\&&
+i\bar A^-_{B12}\Bigl[
(\bar A^-_{B12}+A^-_{F11})(\bar A^+_{B12}+A^+_{F11})
-A_{F12}^-A^+_{F12}\Bigr]\lambda_+\lambda_-^2 \, ,
\non\\
\non\\
\tilde J_{\rm spin}^{{\otimes}\perp} &=& 
{2\over s}\Bigl[
(S_{B12}^+S_{B12}^- - S_{F12}^+S_{F12}^-)z_-
-(\bar A_{B12}^-+A_{F11}^-)z_+\Bigl(A_{B12}^+ +{1\over z_+}\Bigr)
+z_+A_{F12}^+A_{F12}^-
\Bigr]\lambda_-
\non\\&&
+i\bar A_{B12}^-\lambda_-
(S^+_{B12}S^a_{B12}-S^+_{F12}S^a_{F12})\lambda_a^2
\nonumber\\&&
-i\bar A^-_{B12}\Bigl[(\bar A^+_{B12}+A^+_{F11})^2-(A^+_{F12})^2\Bigr]
\lambda_+^2\lambda_- \nonumber\\&&
-i\bar A^+_{B12}\Bigl[
(\bar A^-_{B12}+A^-_{F11})(\bar A^+_{B12}+A^+_{F11})
-A_{F12}^-A^+_{F12}\Bigr]\lambda_+^2\lambda_- \, .
\non\\
\label{decompJspin}
\ear
The remaining components are obtained using the symmetry

\bear
\tilde J_{\rm scal,spin}^{\oplus\parallel} &=& - \tilde J_{\rm scal, spin}^{\oplus\perp} 
(+\leftrightarrow -) \, ,\non\\
\tilde J_{\rm scal,spin}^{{\otimes}\parallel} &=& \tilde J_{\rm scal, spin}^{{\otimes}\perp}
(+ \leftrightarrow -) \, .\non\\
\label{relJpm}
\ear
The integrands are written in terms of the standard worldline functions
(see (3.21) and (4.9) of part 1)

\bear
S_{B12}^{\pm} &=&
{\sinh(z_{\pm}(1-2u))\over \sinh(z_{\pm})} ,
\non\\
A_{B12}^{\pm} &=&
{\cosh(z_{\pm} (1-2u))\over 
\sinh(z_{\pm})}-{1\over z_{\pm}} ,\nonumber\\
A_{B11}^{\pm} = A_{B22}^{\pm} &=&
\coth(z_{\pm})-{1\over z_{\pm}}, \non\\ 
\bar A_{B12}^{\pm}&=& A_{B12}^{\pm}-A_{B11}^{\pm}
= {\cosh(z_{\pm} (1-2u))-\cosh(z_{\pm})\over 
\sinh(z_{\pm})},\non\\
S_{F12}^{\pm} &=& 
{\cosh(z_{\pm}(1-2u))\over \cosh(z_{\pm})} ,
\non\\
A_{F12}^{\pm} &=& 
{\sinh(z_{\pm} (1-2u))\over 
\cosh(z_{\pm})}  ,\non\\
A_{F11}^{\pm} = A_{F22}^{\pm} &=& {\tanh}(z_{\pm}) \, .\non\\
\label{defSA}
\ear
Note that, as usual in this formalism \cite{vv,review}, the scalar loop integrands 
(\ref{decompJscal}) are obtained from the spinor loop ones (\ref{decompJspin}) simply
by nullifying all fermionic worldline correlators $S^{\pm}_{Fij},A^{\pm}_{Fij}$.

\vspace{10pt}
\section{On-shell amplitudes}
\label{onshell}
\renewcommand{\theequation}{3.\arabic{equation}}
\setcounter{equation}{0}

In vacuum at this point we would use the dispersion relation

\bear
k^2 &=& 0 \, .
\label{dispfree}
\ear
The modifications of this relation due to gravitational corrections are not
relevant for our present purposes, since they would produce terms of higher 
order in $\kappa$. 
However, this is less clear for the field-induced corrections
to the electromagnetic $\Pi^{\mu,\nu}$. 
The question of under which conditions (\ref{dispmod})
can still be well-approximated by (\ref{dispfree}) was, for the magnetic case,
studied in \cite{ditgiebook}. There it was shown that this is the case at least for moderate fields and frequencies, $B \leq O(B_{\rm cr})$ and
$\omega \leq O(m)$, where $B_{cr}={m^2\over e}$ denotes 
the ``critical'' magnetic field
strength ($B_{cr}= 4.4 \times 10^9\, T$ for electrons). 
Here the restriction on the photon frequency is not very significant,
since for frequencies beyond the pair creation threshold $\omega = 2m$
processes involving electron-positron pair creation 
become possible, and then are usually physically more
relevant than the dispersive processes which we are concerned with here.
To the contrary, the bound on the field strength may pose a restriction for
applications to magnetars which are believed to carry field strengths 
up to several orders of magnitude higher than $B_{cr}$ \cite{duntho}.    

In the following, we will assume that the use of $k^2=0$ is justified, and
work out the consequences. Using $k^0=\omega = \vert {\vec k} \vert$ 
eqns. (\ref{lambdaspecial}),(\ref{polspecial}) simplify to
(with $k^1=0$)

\bear
\varepsilon_{\perp}^{\mu} &=& (0,1,0,0)\, , \nonumber\\
\varepsilon_{\parallel}^{\mu} &=& {1\over \sin \theta}
(\cos \theta,0,0,1 ) \, , \nonumber\\
\label{polveryspecial}
\ear
and

\bear
\lambda_{\perp} &=& \omega \sin\theta \, ,\nonumber\\
\lambda_{\parallel} &=& -i\omega \sin\theta \, ,\nonumber\\
\label{lambdaveryspecial}
\ear
where $\theta$ is the angle between the $z$-axis (the field direction) and the direction of 
the photon propagation. 
Since in this approximation the amplitude depends on $\omega$ and $\theta$
only in the combination $\omega \sin\theta$, there is no point in keeping the
dependence on $\theta$. We will therefore restrict ourselves in the following
to the case $\sin\theta = 1$ (propagation perpendicular to the field direction).
For a general field we are then left with the four parameters 
$m$, $\omega$, $B$, and $E$. It will be convenient to work with the three
dimensionless variables 

\bear
\hat\omega  &\equiv & {\omega\over m} \, , \non\\
\hat B &\equiv & {eB\over m^2} = {B\over B_{cr}} \, ,\non\\
\hat E &\equiv & {eE\over m^2} = {E\over E_{cr}} \, .\non\\
\label{defhatoBE}
\ear
Here $E_{cr}={m^2\over e}$ denotes the
electric ``critical'' field strength ($1.3 \times 10^{18} V/m$ 
for electrons). 
Similarly, for the  calculation of the integrals it will be useful
to change to the dimensionless
proper-time variable $\hat s\equiv m^2s$. 
Moreover, as usual in this type of calculations
we will change from $u$ to $v \equiv 1-2u$.

Finally, since we wish to compare the one-loop
and the tree level contributions, we normalize the former by the latter.
Thus we will have to compute

\bear
\hat \Pi^{Aa}_{\rm scal,spin}
(\hat\omega,\hat B,\hat E) &\equiv& 
{\rm Re}\Biggl({\bar \Pi^{Aa}_{\rm scal, spin}\over -{i\over 2}\kappa C^{Aa}}\Biggr)
\non\\
\non\\
&=& \alpha \,{\rm Re} \int_0^{\infty}{d\hat s\over \hat s}\,\e^{-i\hat s}\int_0^1dv\,
\hat \pi^{Aa}_{\rm scal,spin}(\hat s,v,\hat\omega,\hat B,\hat E)
\non\\
\label{defPihat}
\ear

\no
with dimensionless integrands $\hat \pi^{Aa}_{\rm scal,spin}$. 
 
\vspace{10pt}
\section{The magnetic field case}
\label{magnetic}
\renewcommand{\theequation}{4.\arabic{equation}}
\setcounter{equation}{0}

We specialize to the purely magnetic case, $E=0$. Then $z_-=0$ so that

\bear
S_{B12}^- &=& v \, ,\non\\
S_{F12}^- &=& 1 \, ,\non\\
A_{B12}^- &=& 0\, ,\non\\
A_{F12}^- &=& 0 \, ,\non\\
{\bar A_{B12}^-\over z_-} &=& \half (v^2-1)\, .\non\\
\label{SAmag}
\ear
Using these identities in
eqs.(\ref{decompJscal}), (\ref{decompJspin}), (\ref{relJpm}) 
one can immediately show that

\bear
\Pi_{\rm scal,spin}^{\oplus\parallel} 
= \Pi_{\rm scal,spin}^{\otimes\perp}
=0 \, . \non\\
\ear
This is in accordance with the CP analysis mentioned above. 

\no
The integrands of the
nonvanishing components are, using (\ref{polveryspecial}), (\ref{lambdaveryspecial}),

\bear
\hat \pi_{\rm scal}^{\oplus\perp}
&=&
{1\over 8\pi}\biggl\lbrace{z_+\over\sinh(z_+)}
\exp\Bigl[z_+\Bigl({\bar A^+_{B12}\over z_+}+\half (1-v^2)\Bigr){\hat\omega^2\over 2\hat B}\Bigr]
\non\\&&\quad\times
\Bigl[
(S_{B12}^+)^2-\bar A^+_{B12}(\bar A^+_{B12}+{1\over z_+})
+\bar A^+_{B12}\Bigl((S_{B12}^+)^2-(\bar A^+_{B12})^2-vS_{B12}^+\Bigr)
{\hat\omega^2\over 2\hat B}\Bigr]
-{2\over 3}
\biggr\rbrace \, ,
\non\\
\hat \pi_{\rm scal}^{\otimes\parallel}
&=&
{1\over 8\pi}\biggl\lbrace{z_+\over\sinh(z_+)}
\exp\Bigl[z_+\Bigl({\bar A^+_{B12}\over z_+}+\half (1-v^2)\Bigr){\hat\omega^2\over 2\hat B}\Bigr]
\non\\&&\quad\times
\Bigl[vS_{B12}^+-{\bar A^+_{B12}\over z_+}
+\bar A^+_{B12}v(S^+_{B12}-v){\hat\omega^2\over 2\hat B}\Bigr]
-{2\over 3}
\biggr\rbrace 
\non\\
\label{hatpimagscal}
\ear 
for the scalar case and

\bear
\hat \pi_{\rm spin}^{\oplus\perp}
&=&
-{1\over 4\pi}\biggl\lbrace{z_+\over\tanh(z_+)}
\exp\Bigl[z_+\Bigl({\bar A^+_{B12}\over z_+}+\half (1-v^2)\Bigr){\hat\omega^2\over 2\hat B}\Bigr]
\non\\&&\quad\times
\Bigl[
(S_{B12}^+)^2 - (S_{F12}^+)^2 + (A_{F12}^+)^2 
-\Bigl(\bar A^+_{B12}+A_{F11}^+\Bigr)\Bigl(\bar A^+_{B12}+{1\over z_+}+A_{F11}^+\Bigr)
\non\\
&&\qquad
+\bar A^+_{B12}\Bigl((S_{B12}^+)^2 - (S_{F12}^+)^2 -(\bar A^+_{B12}+A_{F11}^+)^2
+(A_{F12}^+)^2-vS_{B12}^+ +S_{F12}^+\Bigr)
{\hat\omega^2\over 2\hat B}\Bigr]\non\\
&&\qquad\quad
+{4\over 3}
\biggr\rbrace \, ,
\non\\
\hat \pi_{\rm spin}^{\otimes\parallel}
&=&
-{1\over 4\pi}\biggl\lbrace{z_+\over\tanh(z_+)}
\exp\Bigl[z_+\Bigl({\bar A^+_{B12}\over z_+}+\half (1-v^2)\Bigr){\hat\omega^2\over 2\hat B}\Bigr]
\non\\&&\quad\times
\biggl[vS_{B12}^+ - S_{F12}^+ -{1\over z_+}\Bigl(\bar A^+_{B12}+A_{F11}^+\Bigr)
+\bar A^+_{B12}\Bigl(vS^+_{B12}-S_{F12}^++1-v^2\Bigr){\hat\omega^2\over 2\hat B}\biggr]
\non\\
&&\qquad\quad
+{4\over 3}
\biggr\rbrace
\non\\
\label{hatpimagspin}
\ear 
for the spinor case.
Here $z_+=i\hat B \hat s$ and
$S_{B12}^+, \bar A_{B12}^+$ etc. are as in (\ref{defSA}) with $1-2u$ replaced by $v$.

The resulting integrals (\ref{defPihat}) have a structure similar to the parameter integrals for
the photon vacuum polarization in a constant magnetic field obtained by Tsai and Erber \cite{tsaerb}.
Evaluating this type of integral for the whole range of field strengths and photon energies is
known to be difficult \cite{tsaerb,melsto,artimovich,kalitz,kohyam}, and in fact appears to
have never been done in full generality for the photon-photon case. 
For energies below the pair creation energy a direct numerical
calculation is unproblematic after rotating to Euclidean proper time, $\hat T \equiv i \hat s$.
From eqs. (\ref{defPihat}),(\ref{hatpimagscal}),(\ref{hatpimagspin}),(\ref{defSA})
it is evident that the $v$ - integral is always finite, and the integrand of the
resulting $\hat T$ - integral is exponentially falling for small $\hat\omega$. 
The pair creation threshold $\hat \omega_{\rm cr}$
can be recognized precisely through the fact that for $\hat \omega > \hat \omega_{\rm cr}$
the $\hat T$ -- integral becomes divergent at large $\hat T$. 
Thus, analyzing the asymptotic behaviour of
the integrands for large $\hat T$ at fixed $B$ one finds that, for the scalar case,
the critical value $\hat\omega_{\rm cr}$ is the same for both polarization components,
while for the spinor case it depends on the polarization:

\bear
\hat\omega_{\rm cr,scal}^{\oplus \perp} =
\hat\omega_{\rm cr,scal}^{\otimes \parallel}  &=& 2\sqrt{1+\hat B}  \, ,
\non\\
\hat\omega_{\rm cr,spin}^{\oplus\perp} &=& 1+\sqrt{1+2\hat B} \, ,
\non\\
\hat\omega_{\rm cr,spin}^{\otimes\parallel} &=& 2 \, .\non\\
\label{omegacrit}
\ear
This divergence signals the onset of real pair creation, although to study the
pair creation process itself one would have to consider the full photon-graviton
polarization matrix (see (\ref{dispbig}) above). 
All four threshold values (\ref{omegacrit}) agree
with what one finds for the corresponding photon-photon amplitudes
\cite{toll,adler,vv,kalitz}. 

Beyond the critical energy Euclidean proper time cannot be used any more,
while a numerical integration in the original proper -- time variable $\hat s$
poses enormous difficulties, due to the combined effect of the 
oscillatory behaviour of the universal exponential factor 
and the poles of the trigonometric prefactor functions. 

Thus we will restrict our investigation to the two regimes which have also  been
well-studied in the photon-photon case, namely (i) photon energies below
the pair creation threshold with arbitrary $B$ 
and (ii) arbitrary photon energies at low field strength.
Special attention will be given to the limit of zero photon energy,
since, as we will see, here one can obtain the amplitudes in closed form. 

\subsection{Photon energies below threshold}
\label{belowthreshold}

\begin{figure}[ht]
\centering
\includegraphics{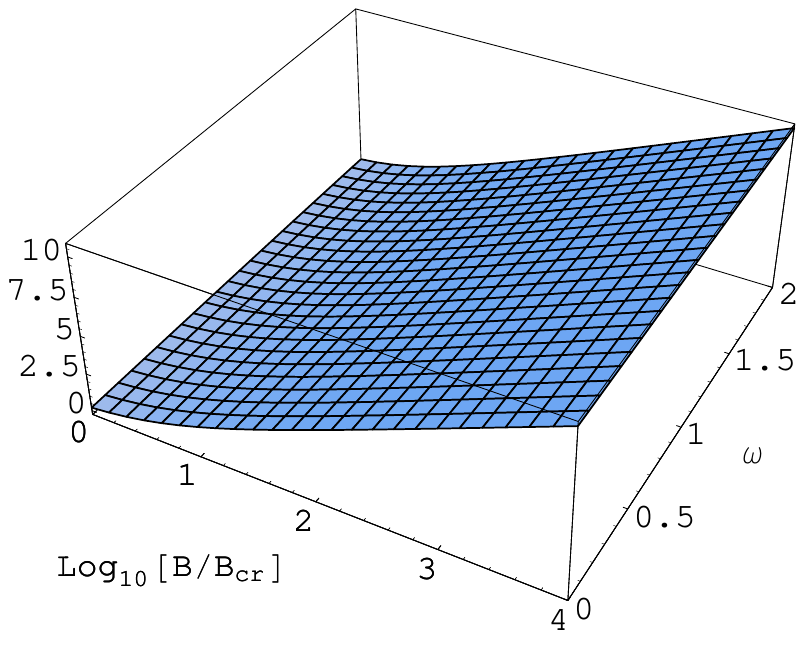}
\caption{Numerical plot of the amplitude $\hat\Pi_{\rm spin}^{\oplus \perp}(\hat\omega,\hat B)$ for 
$\hat B$ between $1$ and $10^4$ and $\hat\omega\leq 2 $. A global factor of 
$-{\alpha \over 4\pi}$ has been omitted.
}
\label{fig2}
\end{figure}

As mentioned above, below the pair creation threshold the parameter integrals 
(\ref{defPihat}) are, after a Wick rotation $\hat T = i\hat s$,
suitable for a direct numerical evaluation at arbitrary
magnetic field strength.
Fig. \ref{fig2} shows the results of such an evaluation, using MATHEMATICA,
for the spinor loop amplitude ratio
$\hat\Pi_{\rm spin}^{\oplus \perp}(\hat\omega,\hat B)$.
The amplitude ratio is shown for field strengths in the range
$B_{\rm cr} < B < 10^4 B_{\rm cr}$ and photon/graviton energies 
$\hat \omega \le 2$. 
A global factor of $-{\alpha\over 4\pi}$ has been omitted.

\medskip

The corresponding plots for $\hat\Pi_{\rm spin}^{\otimes\parallel}$
and $\hat \Pi^{\oplus\perp}_{\rm scal},\hat \Pi^{\otimes\parallel}_{\rm scal}$ 
are very similiar for photon/graviton
energies below the corresponding thresholds 
(the numerical integration becomes unstable 
for $\omega$ too close to $\omega_{\rm crit}$). 
In particular, all four amplitude ratios display a logarithmic growth
in $B$ for large $B \gg B_{\rm cr}$. And indeed, 
for fixed $\omega <  \omega_{\rm cr}$ and $\hat B\to\infty$ it is easy to show
the following asymptotic behaviour of 
(the nonvanishing components of)
$\hat \Pi^{Aa}_{\rm scal,spin}
(\hat\omega,\hat B)$,

\bear
\hat \Pi^{Aa}_{\rm scal}
(\hat\omega,\hat B) &\stackrel{\hat B\to\infty}{\sim}&
-{\alpha\over 12\pi}\ln (\hat B) \, ,
 \non\\ 
\hat \Pi^{Aa}_{\rm spin} 
(\hat\omega,\hat B) &\stackrel{\hat B\to\infty}{\sim}& 
-{\alpha\over 3\pi}\ln (\hat B) \, . \non\\ 
\label{pihatlargeB}
\ear
Thus this asymptotic behaviour 
is independent of the photon energy, as well as of the 
polarization choice. Moreover, it is easy to see that it relates directly to
the renormalization terms in the integrands (\ref{hatpimagscal}),(\ref{hatpimagspin}).
Thus we recognize here a connection between the short-distance behaviour and the
strong-field limit of this amplitude which is familiar from the photon-photon case,
as well as other external field
processes \cite{ritus_connection,dugisc}. 

\subsection{Zero photon energy at arbitrary magnetic field strength}
\label{lowphotonmag}

Closed-form results can be obtained in the zero energy limit.
Setting $\hat \omega =0$ in (\ref{hatpimagscal}),(\ref{hatpimagspin}) one finds
that the $v$ integrals can be done analytically, yielding trigonometric functions
of $z_+$:

\bear
\int_0^1dv \,\hat\pi^{\oplus\perp}_{\rm scal}
&=& 
{1\over 8\pi}
\biggl\lbrack 
-{3\over 2}z_+\,{\cosech}^3 z_+ - {z_+^2{\coth}^2z_+ - 6z_+ {\coth} z_+ + z_+^2  + 2 \over 2z_+ 
{\sinh} z_+} -{2\over 3} 
\biggr\rbrack \, ,\non\\
\int_0^1dv \,\hat\pi^{\otimes\parallel}_{\rm scal}
&=& {1\over 8\pi}
\biggl\lbrack 
{2\cosh z_+ \over \sinh^2 z_+} -{2\over {z_+\sinh z_+}} -{2\over 3}\biggr\rbrack \, ,\non\\
\int_0^1dv \,\hat\pi^{\oplus\perp}_{\rm spin}
&=&
-{1\over 4\pi}
\biggl\lbrack 
3\,{\cosech}^2 z_+ - {\coth} \, z_+ \Bigl(2z_+\,{\cosech}^2\,z_+ + {1\over z_+}\Bigr) + {4\over 3}
\biggr\rbrack \, ,
\non\\
\int_0^1dv \,\hat\pi^{\otimes\parallel}_{\rm spin}
&=&
-{1\over 4\pi}
\biggl\lbrack 
2\,{\cosech}^2z_+ - 2{{\coth} \, z_+\over z_+} +{4\over 3}
\biggr\rbrack \, .
\non\\
\label{lowphotv}
\ear
Thus the remaining proper-time integrals are of the same type as the standard
proper-time representations of the magnetic Euler-Heisenberg Lagrangians 
(see, e.g., \cite{geraldreview}),

\bear
{\cal L}_{\rm scal}^{\rm EH}(\hat B)
&=&
-{m^4\over 16\pi^2} 
\int_0^{\infty}
{d\hat s\over \hat s^3}\,\e^{-i\hat s}
\biggl[{\hat B\hat s\over\sin (\hat B\hat s)} - {(\hat B\hat s)^2\over 6} -1 \biggr]
\qquad ({\rm Scalar \,\, QED}) \non\\
{\cal L}_{\rm spin}^{\rm EH}(\hat B)
&=&
\,\,\,\,\,
{m^4\over 8\pi^2} 
\int_0^{\infty}
{d\hat s\over \hat s^3}\,\e^{-i\hat s}
\biggl[{\hat B\hat s\over{\tan} (\hat B\hat s)} + {(\hat B\hat s)^2\over 3} -1 \biggr]
\qquad ({\rm Spinor \,\, QED}) \non\\
\label{EHL}
\ear
In fact, they can be expressed in terms of the derivatives of these
Lagrangians as follows:

\bear
\hat\Pi_{\rm scal,spin}^{\oplus \perp}(\hat\omega=0,\hat B) 
&=& 
-{2\pi\alpha\over m^4} \Bigl({1\over \hat B}{\pa\over\pa \hat B} + {\pa^2\over \pa \hat B^2}\Bigr)
\,{\cal L}_{\rm scal,spin}^{\rm EH}(\hat B) \, ,
\non\\
\hat\Pi_{\rm scal,spin}^{\otimes \parallel}(\hat\omega=0,\hat B)
&=& 
-{4\pi \alpha\over m^4}
{1\over \hat B}{\pa\over\pa \hat B}\, {\cal L}_{\rm scal,spin}^{\rm EH}(\hat B) \, .
\non\\
\label{EHtoamp}
\ear
The existence of this connection between the photon-graviton
amplitudes and the Euler-Heisenberg Lagrangians is not an accident.
Eqs. (\ref{EHtoamp}) have been obtained before by Gies and Shaisultanov \cite{giesha}
using the linear coupling of the graviton to the Maxwell stress tensor 
(see eq. (\ref{2})), and exploiting the relation between the vacuum expectation value of
the Maxwell tensor and the effective action,

\bear
\vev{T^\mn} &=& {2\over \sqrt{-g}}\,{\delta \Gamma\over \delta g_\mn} \, .
\ear
Applying (\ref{EHtoamp}) to the explicit representations of the
Euler-Heisenberg Lagrangians in terms of the Hurwitz $\zeta$ - function
\cite{dittrich,ditreubook,blviwi,geraldreview} 

\bear
{\cal L}_{\rm scal}^{\rm EH}(B)
&=&
-{(eB)^2\over 4\pi^2}
\biggl\lbrace
\zeta_H'\Bigl(-1,\half + {m^2\over 2eB}\Bigr)
+ \zeta_H\Bigl(-1,\half + {m^2\over 2eB}\Bigr)\biggl\lbrack 1
+ \ln \Bigl({m^2\over 2eB}\Bigr)\biggl\rbrack
\non\\&& \hspace{45pt}
+{3\over 4} \Bigl({m^2\over 2eB}\Bigr)^2
\biggr\rbrace \, ,
\non\\
{\cal L}_{\rm spin}^{\rm EH}(B)
&=& \!\!
{(eB)^2\over 2\pi^2}
\biggl\lbrace
\zeta_H'\Bigl(-1,{m^2\over 2eB}\Bigr)
+ \zeta_H\Bigl(-1,{m^2\over 2eB}\Bigr)\,\ln \Bigl({m^2\over 2eB}\Bigr)
-{1\over 12}
+\fourth \Bigl({m^2\over 2eB}\Bigr)^2
\biggr\rbrace
\non\\
\label{EHurwitz}
\ear
(the prime on $\zeta_H$ refers to a derivative in the first variable)
one obtains 

\def\gam{{\half + {1\over 2\hat B}}}
\bear
\hat\Pi_{\rm scal}^{\oplus \perp}(0,\hat B) 
&=& 
{\alpha\over 8\pi}
\biggl\lbrace
16 \zeta_H'\Bigl(-1,\gam\Bigr)
+{1\over \hat B^2}\psi\Bigl(\gam\Bigr)
- {6\over \hat B} \ln \Gamma \Bigl(\gam\Bigr)
\non\\&& \hspace{25pt}
+{3\over\hat B} \ln (2\pi) -{2\over 3}\ln (2\hat B)
-{2\over \hat B^2} 
\biggr\rbrace \, ,
\non\\
\hat\Pi_{\rm scal}^{\otimes \parallel}(0,\hat B)
&=& 
{\alpha\over 8\pi}
\biggl\lbrace
16\zeta_H'\Bigl(-1,\half + {1\over 2 \hat B}\Bigr)
-{4\over \hat B}\ln\Gamma \Bigl(\half + {1\over 2 \hat B}\Bigr)
+{2\over \hat B}\ln (2\pi)
\non\\&&\hspace{25pt}
 -{2\over 3}\ln (2\hat B) - {1\over \hat B^2}
+ {1\over 3}
\biggr\rbrace \, ,
\non\\
\hat\Pi_{\rm spin}^{\oplus \perp}(0,\hat B) 
&=& 
-{\alpha\over 4\pi}
\biggl\lbrace
16\zeta_H'\Bigl(-1,{1\over 2 \hat B}\Bigr)
-{6\over \hat B}\ln\Gamma \Bigl({1\over 2 \hat B}\Bigr)
+{1\over\hat B^2}\psi\Bigl({1\over 2\hat B}\Bigr)
\non\\&&\hspace{25pt}
+{1\over\hat B}\ln \Bigl({4\pi^3\over \hat B}\Bigr)
+{4\over 3}\ln (2\hat B) -{2\over\hat B^2} + {1\over \hat B}
\biggr\rbrace \, ,
\non\\
\hat\Pi_{\rm spin}^{\otimes \parallel}(0,\hat B)
&=& 
-{\alpha\over 4\pi}
\biggl\lbrace
16\zeta_H'\Bigl(-1,{1\over 2 \hat B}\Bigr)
-{4\over \hat B}\ln\Gamma \Bigl({1\over 2 \hat B}\Bigr)
 -{2\over\hat B}\ln (\hat B/\pi)
\non\\&&\hspace{25pt}
+ {4\over 3}\ln (2\hat B)
-{1\over\hat B^2}
- {2\over 3}
\biggr\rbrace \, .
\non\\
\label{lowphotmag}
\ear
Here $\psi(x) = \Gamma'(x)/\Gamma(x)$ is the digamma function.

\subsection{Weak magnetic field and arbitrary photon energy}
\label{weakfieldmag}

We proceed to the case of arbitrary photon energy but weak magnetic field, i.e.,
$B \ll B_{\rm cr}$. Our treatment of this case parallels the one 
introduced by Tsai and Erber for the photon-photon case
\cite{tsaerb} (see also \cite{ditgiebook}). 

Assuming $\hat B \ll 1$, we can expand the trigonometric functions appearing
in the integrands (\ref{hatpimagscal}), (\ref{hatpimagspin}), in the
common exponential factor (which is the same as in the photon-photon case \cite{vv})
as well as in the prefactor functions, keeping only the terms of lowest order
in $z_+$. However, since we do not wish to make any assumption on the ratio
$\hat \omega^2\over \hat B$, this truncation has to be done separately
for the terms with and without a factor of $\hat \omega^2\over \hat B$.
For example, the prefactor of $\hat \pi_{\rm scal}^{\oplus\perp}$
involves (see eq. (\ref{hatpimagscal}))

\bear
&&\hspace{-30pt} (S_{B12}^+)^2-\bar A^+_{B12}(\bar A^+_{B12}+{1\over z_+})
+\bar A^+_{B12}\Bigl((S_{B12}^+)^2-(\bar A^+_{B12})^2-vS_{B12}^+\Bigr)
{\hat\omega^2\over 2\hat B} \non\\
&=& \half + {v^2\over 2} + O(z_+^2) + {\hat \omega^2\over\hat B}
\Bigl\lbrack {1\over 48} (v^2-1)^2(3-v^2)z_+^3 + O(z_+^5) \Bigr] \, .
\label{exppimag}\ear
Although the first term in the square bracket is $O(z_+^3)$, at this stage it cannot
be neglected over the leading $\half + {v^2\over 2} $ term.
Similarly, the expansion of the universal exponent in
eqs.(\ref{hatpimagscal}), (\ref{hatpimagspin}) yields

\bear
z_+\Bigl({\bar A^+_{B12}\over z_+}+\half (1-v^2)\Bigr)
\, {\hat\omega^2\over 2\hat B}
&=&
{1\over 48}(1-v^2)^2{\hat\omega^2\over\hat B}z_+^3
+ O(z_+^5) \, .
\ear
Here the leading order term, although of order $O(z_+^3)$, cannot be neglected with respect to the 
exponent of the global factor $\e^{-i\hat s}$.

After performing these truncations, the amplitude ratios turn out to
depend on $\hat B$ and $\hat \omega$
only in the combination $\hat B \hat \omega$. This motivates the introduction
of a new parameter $\lambda$,

\bear
\lambda &\equiv& {3\over 2}\hat B \hat \omega \, .
\label{defotherlambda}
\ear
Moreover, for the following it will be useful to 
interchange the orders of integrations in (\ref{defPihat}),
and perform a $v$ - dependent change of variables 
of the global proper-time variable from $\hat s$ to $x$,

\bear
\hat s &=& \rho x 
\label{shattox}
\ear
where

\bear
\rho &=& \Bigl({6\over \lambda (1-v^2)}\Bigr)^{2\over 3} \, .
\label{defrho}
\ear
After this truncation and change of variables, 
the amplitude ratios take the following form:

\bear
\hat\Pi_{\rm scal}^{\oplus \perp}(\lambda)
&=&
{\alpha\over 8 \pi}\, {\rm Re} \int_0^1 dv \int_0^{\infty}{dx\over x} \non\\
&& \times \Biggl\lbrack \half\Bigl(v^2+1\Bigr)(\e^{-i\Xi}-\e^{-i\rho x})
- {i\over 3} (3-v^2)x^3\e^{-i\Xi}\Biggr\rbrack \, ,\non\\
\hat\Pi_{\rm scal}^{\otimes \parallel}(\lambda)
&=&
{\alpha\over 8 \pi}\, {\rm Re} \int_0^1 dv \int_0^{\infty}{dx\over x} \non\\
&& \times \Biggl\lbrack \half\Bigl(v^2+1\Bigr)(\e^{-i\Xi}-\e^{-i\rho x})
-i {2\over 3} v^2x^3\e^{-i\Xi}\Biggr\rbrack \, ,\non\\
\hat\Pi_{\rm spin}^{\oplus \perp}(\lambda)
&=&
-{\alpha\over 4 \pi}\, {\rm Re} \int_0^1 dv \int_0^{\infty}{dx\over x} \non\\
&& \times \Biggl\lbrack \half\Bigl(v^2 -3\Bigr)(\e^{-i\Xi}-\e^{-i\rho x})
+ {i\over 3} (3+v^2)x^3\e^{-i\Xi}\Biggr\rbrack \, ,\non\\
\hat\Pi_{\rm spin}^{\otimes \parallel}(\lambda)
&=&
-{\alpha\over 4 \pi}\, {\rm Re} \int_0^1 dv \int_0^{\infty}{dx\over x} \non\\
&& \times \Biggl\lbrack \half\Bigl(v^2 -3\Bigr)(\e^{-i\Xi}-\e^{-i\rho x})
+ i{2\over 3} (3-v^2)x^3\e^{-i\Xi}\Biggr\rbrack  \, .\non\\
\label{Pihatapprox}
\ear
The common exponential factor $\e^{-i\Xi}$ is now of the Airy form, 

\bear
\Xi &=& \rho x + \third x^3 \, .
\label{Xinew}
\ear
Note that we have also rewritten the counterterms in a way which will be convenient in 
the following.

Now, the two different $x$ - integrals appearing here can both be expressed in terms
of the modified Airy function $Gi(\rho)$, defined by

\bear
Gi(\rho) &\equiv & {1\over\pi} \int_0^{\infty} dx \sin \Bigl(\rho x + \third x^3\Bigr) \, .
\label{defGi}
\ear
Namely, one has

\bear
f_1(\rho) &\equiv & 
\int_0^{\infty}{dx\over x} 
\Bigl\lbrack \cos \Bigl(\rho x + \third x^3\Bigr) - \cos (\rho x) \Bigr\rbrack 
\non\\
&=& 
\int_0^{\infty}{dx\over x}\biggl[ \cos \Bigl(\third x^3\Bigr)-x\int_0^{\rho}d\rho'\,\sin
\Bigl(\rho' x + \third x^3\Bigr) - \cos (\rho x) \biggr\rbrack  \non\\
&=&
-\pi \int_0^{\rho}d\rho' \, Gi(\rho') + \ln \rho +{2\over 3}\gamma + \third \ln 3,
\non\\
\label{idairy1}
\ear
and

\bear
f_2(\rho) &\equiv& \int_0^{\infty} dx \, x^2 \sin \Bigl(\rho x + \third x^3\Bigr) 
=-\pi\, Gi''(\rho)  = 1 -\pi\rho\, Gi(\rho) \, .\non\\
\label{idairy2}
\ear
This brings us to our final result,

\bear
\hat\Pi_{\rm scal}^{\oplus \perp}(\lambda)
&=&
{\alpha\over 8 \pi} \int_0^1 dv 
\Biggl\lbrack \half\Bigl(v^2+1\Bigr)f_1(\rho)
- {1\over 3} (3-v^2)f_2(\rho)\Biggr\rbrack \, , \non\\
\hat\Pi_{\rm scal}^{\otimes \parallel}(\lambda)
&=&
{\alpha\over 8 \pi} \int_0^1 dv 
\Biggl\lbrack \half\Bigl(v^2+1\Bigr)f_1(\rho)
- {2\over 3} v^2f_2(\rho)\Biggr\rbrack \, ,\non\\
\hat\Pi_{\rm spin}^{\oplus \perp}(\lambda)
&=&
-{\alpha\over 4 \pi} \int_0^1 dv
 \Biggl\lbrack \half\Bigl(v^2 -3\Bigr)f_1(\rho)
+ {1\over 3} (3+v^2)f_2(\rho)\Biggr\rbrack \, ,\non\\
\hat\Pi_{\rm spin}^{\otimes \parallel}(\lambda)
&=&
-{\alpha\over 4 \pi} \int_0^1 dv 
\Biggl\lbrack \half\Bigl(v^2 -3\Bigr)f_1(\rho)
+ {2\over 3} (3-v^2)f_2(\rho)\Biggr\rbrack \, .\non\\
\label{Pihatapproxfin}
\ear
Using (\ref{idairy1}), (\ref{idairy2}) these integrals can be done
numerically without difficulties for any value of $\lambda$.

As in the photon-photon case \cite{tsaerb}, exact results can be obtained in 
the limits of small and large $\lambda$. Using the known asymptotic properties
of the function $Gi(x)$ \cite{olver,nikrit_airy} it is easy
to show that, for small $\rho$,  

\bear
f_1(\rho) &\sim& \ln (\rho) + {2\over 3}\gamma + {1\over 3} \ln 3 + O(\rho) \, , \non\\
f_2(\rho) &=& 1 + O (\rho) \, ,\non\\
\label{fsmallrho}
\ear 
while for $\rho \to\infty$

\vspace{-20pt}

\begin{figure}[h]
\centering
\includegraphics{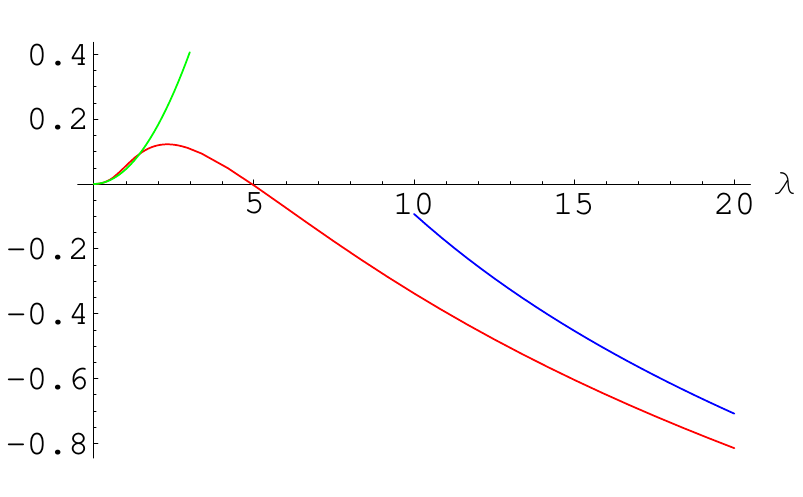}
\caption{Numerical plot of the amplitude $\hat\Pi_{\rm spin}^{\oplus \perp}(\lambda)$ for 
$\lambda$ between $0$ and $20$. The small and large $\lambda$ approximations 
are also shown.
A global factor of  ${\alpha \over 4\pi}$ has been omitted.}
\label{fig3}
\end{figure}

\bear
f_1(\rho) &\sim& {2\over 3} {1\over\rho^3} + O\Bigl({1\over \rho^6}\Bigr) \, ,\non\\
f_2(\rho) &\sim& -{2\over \rho^3} + O\Bigl({1\over \rho^6}\Bigr) \, .\non\\
\label{flargerho}
\ear
Using the leading terms of the expansions (\ref{flargerho})
in (\ref{Pihatapproxfin}) yields the following results for the leading order
terms of the amplitude ratios in the small $\lambda$ limit:

\bear
\hat\Pi_{\rm scal}^{\oplus \perp}(\lambda)
&=&
{\alpha\over 8 \pi} {32\over 945}\,\lambda^2 + O(\lambda^4) \, ,
\non\\
\hat\Pi_{\rm scal}^{\otimes \parallel}(\lambda)
&=&
{\alpha\over 8 \pi}
{8\over 945}\,\lambda^2 + O(\lambda^4) \, ,
\non\\
\hat\Pi_{\rm spin}^{\oplus \perp}(\lambda)
&=&
{\alpha\over 4 \pi} 
{128\over 2835}\, \lambda^2 + O(\lambda^4) \, ,
\non\\
\hat\Pi_{\rm spin}^{\otimes \parallel}(\lambda)
&=&
{\alpha\over 4 \pi}
{40\over 567} \,\lambda^2 + O(\lambda^4) \, .
\non\\
\label{Pihatsmalllambda}
\ear
Using the leading terms of the expansions (\ref{fsmallrho})
gives the asymptotic behaviour for large $\lambda$:

\bear
\hat\Pi_{\rm scal}^{\oplus \perp}(\lambda)
&= & 
{\alpha\over 8 \pi} 
\biggl\lbrack
-{4\over 9}\ln(\lambda) + 
{4\over 9}\gamma + {2\over 3}\ln 3 - {4\over 9}\ln 2 + {2\over 27}
+ O(\lambda^{-{2\over 3}}) 
\biggr\rbrack \, ,\non\\
\hat\Pi_{\rm scal}^{\otimes\parallel}(\lambda)
&= & 
{\alpha\over 8 \pi} 
\biggl\lbrack
-{4\over 9}\ln(\lambda) + {4\over 9}\gamma + {2\over 3}\ln 3 - {4\over 9}\ln 2 
+ {20\over 27}
+ O(\lambda^{-{2\over 3}}) 
\biggr\rbrack \, ,\non\\
\hat\Pi_{\rm spin}^{\oplus \perp}(\lambda)
&=&
{\alpha\over 4 \pi} 
\biggl\lbrack
-{8\over 9}\ln(\lambda) + 
{8\over 9}\gamma + {4\over 3}\ln 3 - {8\over 9}\ln 2 
+ {16\over 27}
+ O(\lambda^{-{2\over 3}}) 
\biggr\rbrack \, ,\non\\
\hat\Pi_{\rm spin}^{\otimes \parallel}(\lambda)
&=&
{\alpha\over 4 \pi} 
\biggl\lbrack
-{8\over 9}\ln(\lambda) + {8\over 9}\gamma + {4\over 3}\ln 3 - {8\over 9}\ln 2 
- {2\over 27}
+ O(\lambda^{-{2\over 3}}) 
\biggr\rbrack \, .\non\\
\label{Pihatlargelambda}
\ear
Note that the leading logarithmic terms in (\ref{Pihatlargelambda})
are independent of the polarization choice.
As in the case of the leading asymptotic growth for large field strength, eq. (\ref{pihatlargeB}),
their coefficients are directly related to the UV counterterms.
This is another fact which is familiar from the photon-photon case.
Fig. \ref{fig3} shows the result of a numerical
evaluation, using MATHEMATICA, of the parameter integral (\ref{Pihatapproxfin}) 
for the amplitude $\hat\Pi_{\rm spin}^{\oplus \perp}(\lambda)$.
The small and large $\lambda$ approximations (\ref{Pihatsmalllambda}), 
(\ref{Pihatlargelambda}) are also shown.

\vspace{10pt}

\vspace{10pt}

\vspace{10pt}
\section{The electric field case}
\label{electric}
\renewcommand{\theequation}{5.\arabic{equation}}
\setcounter{equation}{0}

The photon-graviton conversion process in a constant
field seems to have been studied hitherto exclusively for the
magnetic field case. The reason is, of course, the absence of
evidence for the existence of very strong macroscopic
electric fields anywhere in nature. 
Still, in the near future it may be possible to reach field strengths of the order of
$E_{\rm cr}$ in the laboratory using optical or X-ray lasers \cite{hein_etal,slac,xfel}.
Thus it seems worthwhile to shortly discuss also the electric case.
Moreover, some of our results for the magnetic case carry over to the electric case
simply by analytic continuation. 
Namely, electric-magnetic duality for this amplitude
takes the following form (see eqs.(\ref{poldecomptree}) and (\ref{relJpm})),

\bear
\hat\Pi^{\oplus\parallel}_{\rm scal,spin}(\hat\omega, 0, \hat E) 
&=& \hat\Pi^{\oplus\perp}_{\rm scal,spin}(\hat\omega,\hat B, 0)
\biggl\vert_{B\to iE,\,\omega\to - i\omega}\non\\
\hat\Pi^{\otimes\perp}_{\rm scal,spin}(\hat\omega, 0, \hat E) 
&=& \hat\Pi^{\otimes\parallel}_{\rm scal,spin}(\hat\omega,\hat B, 0)
\biggl\vert_{B\to iE,\,\omega\to - i\omega}\non\\
\label{magtoelec}
\ear
Despite of this formal duality, there is a structural difference between the electric and
magnetic amplitudes. While the magnetic amplitude ratios are real for photon energies
below the pair creation threshold, the integrands of the corresponding electric ones have poles 
in the Euclidean proper-time variable $T=is$, indicating the existence of an imaginary part for
all photon energies. This again agrees with the
photon-photon case \cite{artimovich}, and physically
corresponds to the fact that both the photon/graviton
and the electric field are capable of pair production, while in the magnetic case
the field alone cannot induce an absorptive part.  
Although we are interested here only in the real parts, technically this implies
that a calculation of the full electric amplitudes through numerical computation
of the parameter integrals (\ref{defPihat}) is even more difficult than for the magnetic
amplitudes. In particular, the rotation to Euclidean proper-time is much less useful here,
even for $\omega$ below threshold, since the rotated integrand has poles, both in the
prefactor functions and in the exponent. 
We will therefore restrict our discussion to 
two limiting cases where we can directly draw on our results for
the magnetic case, the zero photon energy case and the weak field case.

\subsection{Zero photon energy and arbitrary electric field strength}
\label{lowphotelec}

For $\omega =0$, in eqs.(\ref{lowphotmag}) we have the magnetic amplitudes in
an explicit form involving only the functions $\Gamma$, $\psi$, and $\zeta_H$.
Implementing the analytic continuation $B\to iE$ is therefore straightforward.
In particular,  for large $E$ the amplitude ratio shows the same logarithmic
growth as we had found in the magnetic case (eq.(\ref{pihatlargeB})),

\bear
\hat \Pi^{Aa}_{\rm scal}
(\hat\omega=0,\hat E) &\stackrel{\hat E\to\infty}{\sim}&
-{\alpha\over 12\pi}\ln (\hat E) \, ,
 \non\\ 
\hat \Pi^{Aa}_{\rm spin} 
(\hat\omega = 0,\hat E) &\stackrel{\hat E\to\infty}{\sim}& 
-{\alpha\over 3\pi}\ln (\hat E)  \, .\non\\ 
\label{pihatlargeE}
\ear

\subsection{Weak electric field and arbitrary photon energy}
\label{weakfieldelec}

Things become even simpler in the weak field limit.
The approximation used in section \ref{weakfieldmag},
based on a weak field expansion and truncation to the leading order
terms, removes the additional poles contained in the electric integrand,
leading to a complete symmetry between the electric and magnetic
cases. Since in this approximation the amplitude ratios depend 
only on the single variable $\lambda = {3\over 2}\hat B \hat \omega$,
eqs.(\ref{magtoelec}) turn into

\bear
\hat\Pi_{\rm scal, spin}^{\oplus \parallel}\Bigl(\lambda = {3\over 2}\hat E \hat \omega\Bigr)
&=&
\hat\Pi_{\rm scal, spin}^{\oplus \perp}\Bigl(\lambda = {3\over 2}\hat B \hat \omega\Bigr) \, ,
\non\\
\hat\Pi_{\rm scal, spin}^{\otimes \perp}\Bigl(\lambda = {3\over 2}\hat E \hat \omega\Bigr)
&=&
\hat\Pi_{\rm scal, spin}^{\otimes \parallel}\Bigl(\lambda = {3\over 2}\hat B \hat \omega\Bigr) \, .
\non\\
\label{magtoeleclambda}
\ear
Therefore all the results of section \ref{weakfieldmag} carry over to the electric
case {\it mutatis mutandis}.

\vspace{10pt}
\section{Conclusions}
\label{conclusions}
\renewcommand{\theequation}{6.\arabic{equation}}
\setcounter{equation}{0}

We have analyzed the parameter integral representations
obtained in part 1 for the one-loop photon-graviton amplitudes in a constant
field at about the same level of detail as was previously achieved for the
analogous QED photon-photon case 
\cite{batsha,tsaerb,melsto,ditgiebook,kohyam,artimovich}. 
In the purely magnetic
case, we have shown that our representation is amenable to a direct
numerical evaluation for photon/graviton energies below threshold,
at arbitrary field strength.  
For weak magnetic fields, a one-parameter integral representation involving
Airy functions has been obtained, and shown to be suitable to
numerical evaluation. Closed-form results have been found for the 
zero energy limit. 
We have also transformed
our results for the weak field and zero energy magnetic cases 
to corresponding results for the electric field case.

The qualitative properties of the photon-graviton amplitudes turn out to
be closely analogous to the ones of the corresponding photon-photon amplitudes.
In particular, in the magnetic case 
they have the same same pair creation thresholds, and a similar asymptotic growth
for strong fields or large photon/graviton energies. This is, of course, not surprising,
particularly considering that the photon-photon and photon-graviton amplitudes are connected
by the gravitational Ward identity ((A.13) of part 1).
Apart from the pair creation thresholds (\ref{omegacrit}),
none of our results show any substantial differences between the scalar and
spinor loop cases. 

From a quantitative point of view, our analysis of the one-loop photon-graviton
amplitude can be summarized as follows. For small photon energies
and large magnetic or electric field strengths, this ratio grows logarithmically
in the field strength (see (\ref{pihatlargeB}) resp. (\ref{pihatlargeE})). 
For weak fields and large photon energies it grows logarithmically
with the photon energy (see (\ref{Pihatlargelambda}), (\ref{magtoeleclambda})).
In these limits it is clearly not possible to compensate the small
prefactor $\alpha/4\pi$ for physically relevant values of the parameters. 
While we have not been able to perform a quantitative analysis in the
whole two-parameter space of field strengths and energies, 
a compound
effect of large field strengths and large energies appears
to be exluded by the fact that the integrand of (\ref{defPihat}) 
contains $\omega$ only in the combination $\hat\omega^2/\hat B$
resp. $\hat\omega^2/\hat E$. This is also borne out by the numerical
results of section \ref{belowthreshold}. Overall,
we conclude that the magnitude of the one-loop contribution to this amplitude will 
not amount to more than a few percent of the tree level one
for physically realistic values of the parameters.

It should be mentioned, though, that there is also a qualitative difference to the
tree-level amplitude. As has been stressed in \cite{cilhar}, the tree level photon-graviton
conversion does, contrary to the photon-axion case, not lead to a dichroism
effect for photon beams. This is because, according to (\ref{poldecomptree}), both
photon polarization components have equal conversion rates.
This symmetry does not extend to the one-loop level (except for the strong field limit). 

As we have seen, in the worldline formalism the calculation of the photon-graviton
polarization tensor in a constant field is only moderately more difficult than the
one of the photon-photon polarization tensor \cite{vv}. 
We expect that even the graviton-graviton
case will be quite feasible. In a future sequel, we intend to analyze this case at the
same level of the photon-graviton one, which then would make it possible to
study the complete set of 
one-loop photon-graviton dispersion relations (\ref{dispbig}).

\vspace{15pt}

\noindent
{\bf Acknowledgements:}
We thank J. Ehlers, H. Gies, G. Raffelt, and S. Theisen  for helpful discussions.
F.B., C.S., and V. V.  thank the Albert-Einstein Institute, Potsdam, for hospitality
during part of this work. 
C. S. and V. V.
thank CONACYT for support through grant PROYECTO 38293-E. 

\begin{appendix}

\section{Quadratic expansion of the Einstein-Maxwell theory}
\label{appward}
\renewcommand{\theequation}{A.\arabic{equation}}
\setcounter{equation}{0}

The Einstein-Maxwell theory is described by  
\bea  
S[g,A] =    
\int d^D x\ \sqrt{g}\, \bigg (  
{1\over \kappa^2 } R - {1\over 4}F_{\mu\nu}F^{\mu\nu}  
\bigg )   
\label{EM}  
\eea  
where the metric $g_{\mu\nu}$ has signature $(-,+,+,\dots, +)$,   
$g= |{\rm det}\, g_{\mu\nu}|$, and $\kappa^2 = 16\pi G_N$.  
The spacetime dimension of interest is $D=4$,
but in this section we may as well keep it arbitrary.
We expand $g_{\mu\nu} = \eta_{\mu\nu}+\kappa h_{\mu\nu}$  
and $A_\mu =\bar A_\mu +a_\mu$.  
Then using the short-hand notation 
$h_\mu \equiv \partial ^\alpha h_{\alpha\mu}$ and   
$h\equiv \eta^{\mu\nu} h_{\mu\nu}$ one obtains 
the following quadratic approximation in the fluctuations 
$(h_{\mu\nu}, a_\mu)$ around the background
$(\eta_{\mu\nu}, \bar A_\mu)$
\bea  
S_{(2)} \eqa  
\int d^D x \, \bigg \{  
{1\over 4}   
( h^{\mu\nu} \square h_{\mu\nu}  
- h\square h + 2 h \partial^\mu h_\mu + 2 h^\mu h_\mu)   
+  
{1\over 2} ( a^\mu \square a_\mu +(\partial_\mu a^\mu)^2 )  
\ccr  
 &+&   
{\kappa\over 2} h_{\mu\nu}\Big (    
\bar F^{\mu\alpha}\bar F^\nu{}_\alpha  
- {1\over 4} \eta^{\mu\nu} \bar F^2
\Big)  
\ccr  
&+&   
\kappa h_{\mu\nu}\Big (   
\bar F^{\mu\alpha} f^\nu{}_\alpha  
- {1\over 4} \eta^{\mu\nu} \bar F^{\alpha\beta} f_{\alpha\beta}   
\Big )  
\ccr  
&- &   
{\kappa^2\over 4} \Big [   
\Big ( {1\over 8} h^2 -{1\over 4} h_{\mu\nu}^2 \Big )\bar F^2  
+ h^{\mu\nu} h^{\alpha\beta}  
\bar F_{\mu\alpha}\bar F_{\nu\beta}   
+ (2 h^{\mu \alpha} h^\nu{}_\alpha -h h^{\mu\nu} )  
\bar F_{\mu\beta}\bar F_\nu{}^\beta\Big ]\bigg \} \non\\
\label{2}
\eea    
In the second line of this expression we recognize the linear coupling 
${\kappa\over 2} h_{\mu\nu} \bar T^{\mu\nu}$ of the graviton $h_{\mu\nu}$ 
with the stress tensor of the background electromagnetic field
 $\bar T^{\mu\nu}=\bar F^{\mu\alpha}\bar F^\nu{}_\alpha  
- {1\over 4} \eta^{\mu\nu} \bar F^2 $.
This tadpole vertex indicates that the nontrivial background 
stress tensor tends to curve the space.
The third line in (\ref{2}) gives instead the tree level graviton-photon mixing
in the electromagnetic background. Using plane waves
\be
h_{\m\n}(x) = \ep_{\m\n}  e^{ikx} \ , \qquad
a_{\a}(x) = \ep_{\a}  e^{ik_2 x} 
\ee
we get for this mixing term the vertex 
\bea
\Delta S_{(2)}  
= (2\pi)^D \delta^D(k+k_2)\, \ep_{\m\n}  \ep_{\a}  \, 
(-i \kappa) \,
\Big [ F^{\m\a}k^\n - (F\cdot k)^\m \eta^{\n\a}
+{1\over 2} \eta^{\m\n} (F\cdot k)^\a \Big ] 
\non\\
\eea
which appears in  the path integrals as $e^{i \Delta S_{(2)} }$
(see eq. (1.2)).

The two-point functions (which we denote by $\Pi$),
in either coordinate or momentum space, 
are contained in $S_{(2)}$ (or in the quadratic part of
the full effective action 
$\Gamma_{(2)}   =S_{(2)}  + \Gamma_{(2)}^{{\rm (1-loop)}}  + \cdots $)  
as follows 
\bea  
S_{(2)} 
 \eqa  
\int d^D x \, \bigg \{ 
{1\over 2}  h_{\mu\nu}(x) \Pi^{\m\n,\lambda\rho}(\partial) h_{\lambda\rho}(x) 
+ {1\over 2} a_\a(x) \Pi^{\a,\b} (\partial) a_{\b}(x) \ccr
&& +  h_{\mu\nu}(x) \Pi^{\m\n,\a}(\partial) a_\a(x) +
{\kappa \over 2}  h_{\mu\nu}(x) \bar T^{\mu\nu}(x) \bigg \}\ccr
\eqa  
\int {d^D k\over (2 \pi)^{D}} \, \bigg \{  
 {1\over 2}h_{\mu\nu}(k) \Pi^{\m\n,\lambda\rho}(k) h_{\lambda\rho}(-k) 
+ {1\over 2} a_\a(k) \Pi^{\a,\b} (k) a_{\b}(-k) \ccr
&& +  h_{\mu\nu}(k) \Pi^{\m\n,\a}(k) a_\a(-k) +
{\kappa\over 2}  h_{\mu\nu}(k) \bar T^{\mu\nu}(-k) \bigg \}
\label{6}
\eea
where the Fourier transform of a field is given by
$h_{\m\n}(x) = \int {d^D k\over (2 \pi)^{D}} \, e^{ikx} h_{\m\n}(k) $, 
its inverse by
$h_{\m\n}(k) = \int d^D x \, e^{-ikx} h_{\m\n}(x) $ 
and $\Pi(k)= \Pi(\partial \to -i k)$.

The equations of motion in term of these two-point functions then
read\footnote{A minus sign in the euclidean formulas
(A.11) of paper I is correctly taken into account by
the Wick rotation, as one can easily check by looking at the 
euclidean tree level Maxwell action
$S= \int d^D x\ {1\over 4}F_{\mu\nu}F^{\mu\nu} $,
which indeed is positive definite in euclidean space (as it should).}
\bea
\delta  a_{\a}(k) : &&  
\Pi^{\a,\b} (k) a_{\b}(-k) + \Pi^{\m\n,\a}(-k) h_{\mu\nu}(-k) =0
\ccr
\delta  h_{\mu\nu}(k) : && 
\Pi^{\m\n,\lambda\rho}(k) h_{\lambda\rho}(-k)+
 \Pi^{\m\n,\a}(k) a_\a(-k) = - {\kappa\over 2} \bar T^{\mu\nu}(-k) \non\\
\label{eom}
\eea
and, in particular, one obtains from (\ref{2}) 
\bea
&& \Pi^{\a,\b}_{\rm tree} (k) = k^\a k^\b -k^2 \eta^{\a\b} \ccr
&&
\Pi^{\m\n,\a}_{\rm tree} (k)  = -{i \kappa\over 2} C^{\m\n,\a}(k) 
\label{Pitree}
\eea
with $C^{\m\n,\a}(k)$ as in (\ref{defCmna}).
Also, one gets
\bea
\Pi^{\m\n,\lambda\rho}_{\rm tree} (k) 
\eqa {k^2\over 2} \Big ( \eta^{\mu\nu}\eta^{\lambda\rho}
- {1\over 2} 
(\eta^{\mu\lambda} \eta^{\nu\rho}
+\eta^{\mu\rho}\eta^{\nu\lambda})
\Big )
- {1\over 2} 
(\eta^{\mu\nu} k^\lambda k^\rho + k^\mu k^\nu \eta^{\lambda\rho})\ccr
&& + {1\over 4} 
(\eta^{\mu\lambda } k^\nu k^\rho +
\eta^{\nu\lambda } k^\mu k^\rho +
\eta^{\mu\rho } k^\nu k^\lambda +
\eta^{\nu\rho } k^\mu k^\lambda) + \bar F\ {\rm terms}.\non\\
\label{Pigravtree}
\eea
Finally, note that a constant $\bar T^{\mu\nu}$ gives 
$\bar T^{\mu\nu}(k)= (2 \pi)^{2}\delta^D(k) \bar T^{\mu\nu}$.

\end{appendix}


%
%


\end{document}